# Rapid Atmospheric Vapor Deposition of $H:In_2O_3$ Transparent Conducting Oxide Thin Films

Xiaoyu Guo[1], Hae-Jun Seok[1], Eilidh L. Quinn[1], Matthew K Sharpe[2], Callum. D. McAleese[2], Yi-Teng Huang[1,3], Xinjuan Li[4], Kexue Li[5], Chia-Yu Chang[1], Yongjie Wang[1], John O'Sullivan[6], Katie L. Moore[5], Caterina Ducati[4], Ruy Sebastian Bonilla[6], Han-Ki Kim[7], Abderrahime Sekkat[1,8], Robert L. Z. Hoye[1,*]

[1]Inorganic Chemistry Laboratory, Department of Chemistry, University of Oxford, South Parks Road, Oxford OX1 3QR, UK.

[2]Surrey Ion Beam Centre, University of Surrey, Guildford, Surrey, GU2 7XH, United Kingdom.

[3]Graduate Institute of Photonics and Optoelectronics and Department of Electrical Engineering, National Taiwan University, Taipei, 10617, Taiwan.

[4]Department of Materials Science and Metallurgy, University of Cambridge, Cambridge, CB3 0FS UK.

[5]Department of Materials, Photon Science Institute, University of Manchester, Engineering Building A, Manchester, United Kingdom, M13 9PL.

[6]Department of Materials, University of Oxford, Oxford, OX1 3PH, UK.

[7]School of Advanced Materials Science and Engineering, Sungkyunkwan University, Suwon, Gyeonggi-do, 16419, South Korea.

[8]Univ. Toulouse, CNRS, Toulouse INP, LGC, Toulouse, France.

* Email: robert.hoye@chem.ox.ac.uk (R. L. Z. H.)

## Abstract

Transparent conducting oxides (TCOs) are essential for the optoelectronics industry, but there is a critical gap in cost-effective methods to rapidly deposit low sheet resistance, high transmittance films without damaging delicate materials, including emerging soft semiconductors like metal-halide perovskites. In this work, atmospheric pressure chemical vapor deposition (AP-CVD) is used to synthesize $H:In_2O_3$ films with 7.20±0.01 $\Omega\ \square^{-1}$ sheet resistance (0.50±0.06 m$\Omega$ cm resistivity) and transmittance up to 89% in the near-infrared (NIR), surpassing commercial sputter-deposited indium tin oxide. The growth rate is 40× higher than atomic layer deposition (ALD), and the AP-CVD films are fully processed under atmospheric conditions at only 140 ºC. Comparison of secondary ion mass spectrometry and time-of-flight elastic recoil detection analysis with changes in carrier concentration indicate that H dopants are introduced from the water oxidant. There is an increase in mobility from 40±10 $cm^2\ V^{-1}\ s^{-1}$ to 160±30 $cm^2\ V^{-1}\ s^{-1}$ when changing from $O_2$ to $H_2O$ as the oxidant, which is attributed to H dopants passivating oxygen vacancies that act as carrier scattering centers. This work establishes AP-CVD as a promising method for manufacturing high figure-of-merit TCOs in a rapid, scalable and cost-effective manner, using mild growth conditions compatible with thermally-sensitive materials.

**Keywords:** $H:In_2O_3$, transparent conducting oxide, atmospheric pressure chemical vapor deposition, thin film manufacturing, high-throughput deposition, near-infrared transmittance

## Introduction

Transparent conducting oxides (TCOs) are vital components in a wide range of technologies, including photovoltaics (PVs), microelectronics, touch screen displays, smart windows and sensors [1-5]. In several next-generation applications, including flexible photovoltaics, organic electronics, and temperature-sensitive optoelectronic devices, deposition under mild, atmospheric conditions is highly desirable [6-9]. Polymer substrates and functional layers made from organic or hybrid materials often impose strict restrictions on deposition temperatures (<150 °C), while vacuum-based processing increases system complexity, can increase capital costs, and limits process flexibility. Indium tin oxide (ITO) is the most widely used TCO commercially, due to its high electrical conductivity and optical transmittance in the visible wavelength range [6-9]. However, although atmospheric-pressure and solution-based routes to depositing ITO have been demonstrated [10-13], these methods generally require processing temperatures of ~300–600 °C, highlighting the difficulty of achieving high electrical performance with low-temperature processing under atmospheric conditions. Developing high-mobility TCOs that can be deposited rapidly at low temperature without vacuum infrastructure, and without requiring harsh oxidizing conditions (*e.g.*, use of plasma), therefore addresses an important materials and processing gap.

All high-performance TCOs achieve low resistivity through a combination of high carrier concentration ($n$) and/or carrier mobility ($\mu$), since $\rho = 1/(ne\mu)$. In ITO, Sn incorporation primarily increases carrier concentration. While effective in lowering resistivity, high carrier densities enhance free-carrier absorption (FCA), particularly in the near-infrared (NIR), and increase ionized impurity scattering, which can limit mobility [14, 15]. This trade-off between conductivity with NIR transmittance and mobility becomes increasingly relevant for applications requiring high transmittance beyond the visible spectrum, such as in telecommunications and night-vision devices [16, 17]. Hydrogen-doped indium oxide ($H:In_2O_3$) provides an appealing alternative, in which hydrogen acts as a shallow donor with minimal perturbation of the conduction band. This enables moderate-to-high carrier densities while preserving high mobility [18-20]. In addition, hydrogen can passivate compensating defects and oxygen-vacancy–related trap states, further reducing carrier scattering [21, 22]. Reported mobilities for $H:In_2O_3$ prepared by ALD or sputtering are typically in the range of ~120–130 $cm^2\ V^{-1}\ s^{-1}$, highlighting its potential as a high-mobility TCO that does not require high carrier density to achieve low resistivity [18, 20, 23]. These intrinsic advantages make $H:In_2O_3$ particularly attractive for broadband transparent electrode applications.

Despite these material benefits, deposition strategies remain a bottleneck. Sputter deposition is currently a well-established route for $H:In_2O_3$ deposition, due to its ability to deliver low resistivity (< 0.3 mΩ cm) with high visible and NIR transmittance (>80%) [24, 25]. Nevertheless, the energetic plasma environment associated with sputtering can be detrimental to soft or defect-sensitive materials, including hybrid perovskites and polymeric substrates. Atomic layer deposition (ALD) has also been widely explored for $H:In_2O_3$-based TCOs due to its precise control over film thickness and quality, even on complex substrates [20, 26, 27]. This is achieved through self-limiting reactions and the sequential injection of gases to a reaction chamber. The thickness increment per precursor/oxidant sequence is termed the growth per cycle (GPC, typically expressed in Å per cycle). Numerous studies have shown that $In_2O_3$ can be grown by ALD using a variety of precursors at relatively mild temperatures (70-250 °C [28]), and the precursors have low kinetic energy, such that they do not mechanically damage the underlying materials [29-31]. Various precursors, including $InCl_3$ [20, 27, 32-39] and cyclopentadienyl indium (InCp) [20, 27, 32-39], have been employed with different oxidants in ALD, such as $H_2O/O_2$, $H_2O_2$, $O_2$ plasma, and $N_2O$, yielding a GPC ranging from 0.06 to 1.7 Å per cycle. This inherently low growth per cycle is characteristic of ALD. When film thicknesses of several hundred nanometers are required to achieve low sheet resistance, deposition times can extend to several hours, limiting throughput [40].

Spatial atomic layer deposition (SALD) has been developed to overcome the throughput bottlenecks of ALD by separating precursor streams in space rather than in time [41-45]. In close-proximity SALD, the substrate is held parallel to the gas manifold, and moves continuously beneath isolated precursor zones separated by inert gas curtains. When the manifold–substrate spacing is sufficiently small (typically <10 μm), gas intermixing is suppressed, and surface reactions remain self-limiting, analogous to conventional ALD, with the inert gas curtains taking the place of the purge steps. The primary speed advantage arises from eliminating vacuum purge steps rather than significantly increasing GPC [46-48]. However, if precursor intermixing occurs due to increased manifold–substrate spacing or insufficient gas separation, gas-phase reactions are enabled and the process transitions towards a CVD-like growth regime. For clarity, we distinguish between growth per cycle (GPC) and growth rate. GPC refers specifically to the thickness increment per self-limiting ALD sequence. In contrast, growth rate (nm $min^{-1}$ or Å $min^{-1}$) describes thickness accumulated per unit time and depends on both GPC and cycle duration. In CVD-like processes where growth is not self-limiting, thickness scales directly with exposure time, and the growth rate is therefore the most meaningful metric for comparing how rapidly depositions are carried out.

Conventional ALD of $H:In_2O_3$ using InCp and $H_2O/O_2$ has been reported to deliver high Hall mobilities of 138 $cm^2\ V^{-1}\ s^{-1}$ after post-deposition crystallization at 150-200 °C in flowing $N_2$ [20]. Nevertheless, the need for relatively high thermal budgets and prolonged post-deposition treatment remains a drawback for applications requiring rapid and low-temperature processing. Promising results for the development of $In_2O_3$ using SALD was recently demonstrated using a trimethylindium (TMI) precursor with a GPC of (1.33 Å per cycle) [49]. In this case, ozone was used as an oxidant to enhance the surface kinetics of the oxidation half-cycle. The deposited thin films achieved a Hall mobility of 54.7 $cm^2\ V^{-1}\ s^{-1}$ at 225 ºC deposition temperature. When integrated as the active layer in flexible thin-film transistors (TFTs), the $In_2O_3$ films demonstrated a good field-effect mobility of 69.8 $cm^2\ V^{-1}\ s^{-1}$ [49]. Despite the promise of this approach for microelectronics, the use of relatively high growth temperatures, as well as the use of ozone may pose challenges for applications requiring milder, softer chemical processes.

In this work, we address these pressing gaps in the manufacture of TCOs by introducing the use of atmospheric pressure chemical vapor deposition (AP-CVD) for these materials. This is comprised of a close-proximity SALD reactor operating in CVD growth mode, and we focus on $H:In_2O_3$ as a test case. For this material, it is critical to control the oxidants used. This is because the choice of oxidant influences the growth rate, concentration of oxygen vacancies, and how the hydrogen dopant is introduced, which collectively dictate the electrical, optical and structural properties of the $H:In_2O_3$ films. Therefore, in this work, we systematically varied the oxidizing environments used in AP-CVD $H:In_2O_3$, comparing $O_2/H_2O$, $N_2/H_2O$, $O_2/N_2$, and $O_2$ in the oxidant channels. Depositions took place under fully atmospheric conditions, and to ensure we develop a mild growth process, we focused on deposition temperatures of 100–140 ºC, and avoided the use of ozone or plasma. By correlating oxidant chemistry with film morphology, defect structure, hydrogen content, and optoelectronic properties, we investigate the mechanism of hydrogen doping and how this dopant affects electronic and optical properties. Furthermore, we benchmark these optoelectronic properties against commercial ITO and previously reported $In_2O_3$ grown with slower vacuum-based deposition methods.

**Results and Discussion**

A side view of the gas manifold used to deposit $H:In_2O_3$ thin films by AP-CVD is schematically illustrated in **Figure 1**a. This reactor design is described in detail in Ref. [50]. In this system, $N_2$ gas is bubbled through solid InCp powder heated at 40 °C, and this carries metal precursor vapors to the central channel of the gas manifold (green channel in Figure 1a labelled 'M'). This channel runs throughout the entire width of the manifold, and the precursor vapors are introduced from both sides. A narrow slit guides the precursors vapors from this header channel vertically down to the substrate oscillating beneath the manifold. Oxidants are introduced through adjacent channels, which are spatially separated from the metal precursor channel by inert gas curtains. Herein, we investigated $H_2O$ vapor and $O_2$ gas as the oxidants, and compared the use of $O_2$ or $N_2$ as the carrier gas for these oxidants, *i.e.*, the four combinations tested were: $O_2/H_2O$, $N_2/H_2O$, $O_2$ and $O_2/N_2$.

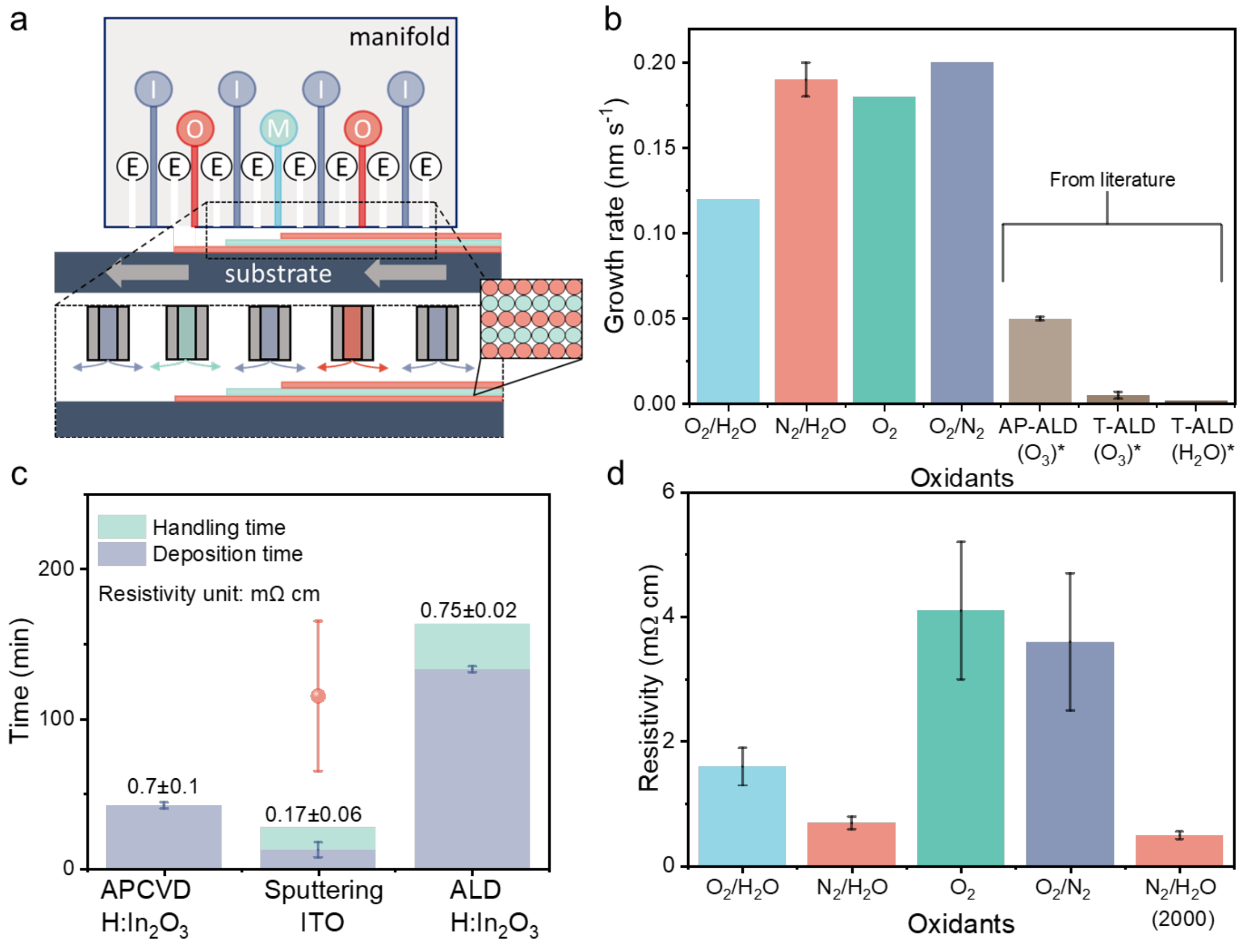

**Figure 1**. $H:In_2O_3$ grown by atmospheric pressure chemical vapor deposition (AP-CVD), compared with established alternatives. **a.** Illustration of the gas manifold used for AP-CVD, "M" denotes the metal precursor, "I" denotes the inert gas ($N_2$), "O" denotes the oxidant gas channels and "E" denotes the exhaust channels. The substrate is loaded onto a platen (metal plate), which heats the sample and oscillates it beneath the gas manifold. **b.** Growth rate of $H:In_2O_3$ thin films deposited by AP-CVD with the four different oxidant combinations tested in this work, compared to the growth rate of $H:In_2O_3$ thin films by conventional ALD reported in the literature [49] (denoted as *). **c.** Total process cycle time to reach comparable bulk resistivity for sputtered ITO, ALD $H:In_2O_3$, and AP-CVD $H:In_2O_3$. Total cycle time includes deposition time and conservative estimates of vacuum handling (pump-down and venting), assumed to be 15 min for sputtering and 30 min for ALD. AP-CVD operates under fully atmospheric conditions and therefore requires no vacuum handling. The red outlier point represents sputtered ITO processes in which a post-deposition heat treatment step is included in the total processing time. Refer to **Table S2**, SI for details. **d.** Electrical properties of $H:In_2O_3$ thin films deposited at 140 °C for 1500 cycles, prepared using the four different oxidant combinations tested herein.

The deposition temperature is set by the heated platen that the substrate is mounted onto, which oscillates beneath the manifold, such that the number of oscillations determines the thickness of the film. In this configuration, one "cycle" corresponds to a single back-and-forth oscillation of the substrate beneath the gas manifold. During each oscillation, the substrate sequentially traverses precursor-rich and oxidant-rich regions. Although it is possible to set this manifold-substrate spacing to <10 µm and achieve ALD growth, in practice, a wider substrate-manifold gap is commonly used because of difficulties in maintaining perfect alignment between the gas manifold and substrate to the micron-level repeatedly across multiple samples. Increasing this spacing to >50 µm allows greater fluctuations in manifold-substrate spacing to be tolerated. However, this large spacing results in gas-phase mixing of the precursors, such that growth is by CVD rather than ALD. Nevertheless, prior work with AP-CVD has shown that similar film properties can be achieved in CVD or ALD growth mode, including high film uniformity, provided that the partial pressure of the precursor gases in their respective channels are the same across the width of the gas manifold, and that the manifold-substrate spacing is uniform [51, 52]. Furthermore, there is the added benefit that CVD results in a higher growth rate than ALD [40, 53]. For these reasons, we herein also used a relatively large manifold-substrate spacing of 100 µm for the growth of $H:In_2O_3$. We checked the growth mode by measuring the film thickness as a function of precursor exposure time (by varying the speed at which the substrate oscillates) and found a linear relationship (**Figure S1**a, SI). Given that we kept the number of oscillations constant at 1500, film thickness is proportional to the GPC. Therefore, the linear relationship between thickness and time, without saturation, is consistent with CVD rather than ALD growth. We therefore refer to this process as atmospheric pressure chemical vapor deposition (AP-CVD) rather than spatial ALD.

We prepared $H:In_2O_3$ films at 140 °C (refer to **Table S1**, SI) for 1500 cycles with the four different oxidant combinations, and estimated the growth rate per unit time from the thicknesses obtained and the total time taken for deposition (42-43 min). This assumes growth rates remain constant throughout the entire deposition process, and therefore provides an estimate of the overall growth rate. As shown in Figure 1b, these growth rates achieved by AP-CVD far exceeded those reported for $H:In_2O_3$ grown by atmospheric pressure conventional ALD (AP-ALD); note that trimethylindium (TMI) was used as the metal precursor for ALD, rather than InCp [49]. Focusing on the $H:In_2O_3$ grown with $N_2/H_2O$ in the oxidant channel, we confirmed the thickness to vary linearly with the number of cycles. The growth rate of 0.19 ± 0.01 nm $s^{-1}$ was 40× higher than that of $H:In_2O_3$ deposited using conventional ALD (~0.005 nm $s^{-1}$) [27].

The high growth rates achievable with AP-CVD directly translate into reduced processing times. Figure 1c presents a comparison of the total processing time required to prepare $H:In_2O_3$ by AP-CVD *vs.* ALD compared to conventional sputter deposition of indium tin oxide (ITO). To make a fair comparison, we determined the processing times required to produce TCO films with similar resistivities (0.2–0.8 mΩ cm, refer to Table S2, SI) rather than the same thickness, since the doping levels in ITO and $H:In_2O_3$ are different. For sputtered ITO, we considered two common industrial workflows: (i) heated-substrate sputtering, where the substrate is held at elevated temperature during deposition (commonly >250–300 °C) to promote *in-situ* film crystallization and optical/electrical quality [54, 55], and (ii) low-temperature deposition followed by post-deposition annealing [56, 57]. In both cases, the deposition time is similar, and on the order of 25 ± 5 min for the film thickness considered, as shown in Table S2, but the effective total cycle time differs because of the annealing/thermal budget. Whilst sputtering benefits from a high deposition rate, the particles deposited would typically have high kinetic energy that could mechanically damage soft organic and hybrid materials (*e.g.*, lead-halide perovskites or organic photovoltaics [51, 58]). By contrast, ALD is a softer deposition method, but, as shown in Figure 1c, requires >5× longer processing time (>160 min) to produce $H:In_2O_3$ with similar resistivity. AP-CVD is capable of bridging the gap between these two methods, combining the ability to deposit films in a soft manner, while also having a short processing time (42 ± 2 min) that is similar to that for sputter deposition. Notably, AP-CVD has negligible handling time, given that processing entirely takes place in atmospheric conditions, without requiring a vacuum chamber, in contrast to sputter deposition or ALD. This comparison highlights the combined advantage of high deposition rate, low deposition temperature and simplified atmospheric operation for AP-CVD compared to the well-established sputter deposition and ALD techniques.

**Table 1**. Comparison of growth rate, resistivity, sheet resistance, carrier density, and mobility of commercial ITO versus $H:In_2O_3$ thin films deposited with various oxidant combinations at 140 ºC. c-ITO was purchased from Kintec Company without any treatment, while all $H:In_2O_3$ were prepared by ourselves for this work by AP-CVD. $H:In_2O_3$ films were deposited with 1500 cycles, unless otherwise specified. The growth per cycle and growth rate data for the c-ITO were not provided by the manufacturer, and so these data are listed here as "not reported" (n.r.). Carrier density uncertainty here was estimated by propagating independent uncertainties in thickness, Hall voltage fitting (typically within 5% [59]), and magnetic field calibration (3% calibration error), using standard root-sum-square error propagation. Mobility uncertainty was obtained by combining resistivity and carrier density uncertainties in quadrature.

| Deposition condition | Thickness (nm) | Growth per cycle (nm) | Growth rate (nm $s^{-1}$) | Resistivity (mΩ cm) | Sheet R (Ω □$^{-1}$) | Carrier density ($10^{19}$ $cm^{-3}$) | Mobility ($cm^2$ $V^{-1}$ $s^{-1}$) |
|---|---|---|---|---|---|---|---|
| c-ITO | 150±3 | n.r. | n.r. | 0.19±0.02 | 13±1 | 122±8 | 27±3 |
| $O_2/H_2O$ | 320±30 | 0.20 | 0.12 | 1.6±0.3 | 64±3 | 5.2±0.6 | 60±10 |
| $N_2/H_2O$ | 470±40 | 0.36±0.01 | 0.19±0.01 | 0.7±0.1 | 14.3±0.5 | 5.7±0.6 | 160±30 |
| $O_2$ | 500±40 | 0.33 | 0.18 | 4±1 | 80±13 | 2.0±0.2 | 80±20 |
| $O_2/N_2$ | 560±50 | 0.37 | 0.20 | 4±1 | 65±12 | 4.1±0.4 | 40±10 |
| $N_2/H_2O$ (2000 cycles) | 710±50 | 0.36±0.01 | 0.19±0.01 | 0.50±0.06 | 7.20±0.01 | 6.4±0.6 | 190±30 |

The electrical properties of the AP-CVD films deposited at 140 °C for 1500 cycles are displayed in Figure 1d and Table 1. Films grown with $N_2/H_2O$ in the oxidant channel exhibited the lowest resistivity amongst all combinations of oxidants investigated, reaching 0.7±0.1 mΩ cm (1500 cycles, 470±40 thickness) with a Hall mobility of 160±30 $cm^2$ $V^{-1}$ $s^{-1}$. With an increase in thickness (2000 cycles, 710±50 thickness), the resistivity decreased further to 0.50±0.06 mΩ cm, while the mobility increased to approach 190±30 $cm^2$ $V^{-1}$ $s^{-1}$. In contrast, using $O_2$ in the oxidant channel ($O_2$ only, or $O_2/N_2$) resulted in higher resistivity (~4 mΩ cm for 1500 cycles), while using $O_2/H_2O$ led to an intermediate resistivity (1.6±0.3 for 1500 cycles). These results indicate that oxidant chemistry strongly influences the electrical properties of these AP-CVD-grown films.

The optical transmittance of $H:In_2O_3$ thin films deposited under different oxidizing environments in the visible region is shown in Figure 2a. Spectra were measured for film-on-glass samples using a bare glass substrate as reference (films thicknesses stated in caption). Given that film transmittance depends on its thickness, caution is required when comparing films of different thicknesses directly. Despite being significantly thicker (470±40 nm) than the $O_2/H_2O$-grown film (320±30 nm), the $N_2/H_2O$-grown film exhibits comparable or higher visible transmittance, indicating favorable intrinsic optical properties. The difference becomes more pronounced in the NIR region (Figure 2b), where the optimized $N_2/H_2O$ film deposited for 2000 cycles (710±50 nm thickness) exhibited an average NIR transmittance of

~89.0%, whereas commercial sputtered ITO (150±3 nm thickness) has only ~67.8% transmittance in the same wavelength range. Unlike ITO, which relies on high carrier concentrations that increase FCA, $H:In_2O_3$ achieves comparable conductivity through higher mobility at more moderate carrier concentrations, thereby suppressing infrared absorption [14, 15]. To further decouple thickness effects, ellipsometry-derived optical constants were examined (Figure S2, SI). In the visible wavelength range (400–700 nm), the extinction coefficient ($k$) remains near zero for both films. In contrast, ITO exhibits a marked increase in $k$ in the NIR, consistent with strong FCA. The $N_2/H_2O$-grown $H:In_2O_3$ film maintains a low extinction coefficient across the NIR region, confirming intrinsically suppressed NIR absorption. The reduced FCA is consistent with achieving low resistivity through high mobility (190±30 $cm^2\ V^{-1}\ s^{-1}$) with moderate carrier density ($6.4\pm0.6 \times 10^{19}\ cm^{-3}$), rather than through extremely high carrier concentrations as in ITO ($1.22\pm0.08 \times 10^{21}\ cm^{-3}$). To elucidate the electronic structure of the best-performing $N_2/H_2O$-grown $H:In_2O_3$ thin film, ultraviolet photoelectron spectroscopy (UPS) was performed to determine the valence band maximum (VBM) as well as work function ($\Phi$) (Figure S3a-b, SI). The optical bandgap, obtained from UV-Vis spectrophotometry (**Figure 2**a), was combined with these values to construct the full band structure (Figure S3c, SI). The Fermi level is positioned close to the conduction band minimum (CBM), indicating degenerate $n$-type behavior consistent with hydrogen-induced shallow donor states.

Finally, we note that whilst the resistivity could be reduced through further increases in film thickness, when we increased the number of cycles of AP-CVD $H:In_2O_3$ beyond 2000 cycles, the transmittance decreased, with only marginal improvements in conductivity (Figure S3, SI). At a deposition temperature of 140 °C, and with $N_2/H_2O$ in the oxidant channels, we found 2000 cycles to strike the optimal balance between conductivity and transmittance. For all subsequent characterization, we focused on the films with thicknesses shown in Table 1, unless otherwise specified.

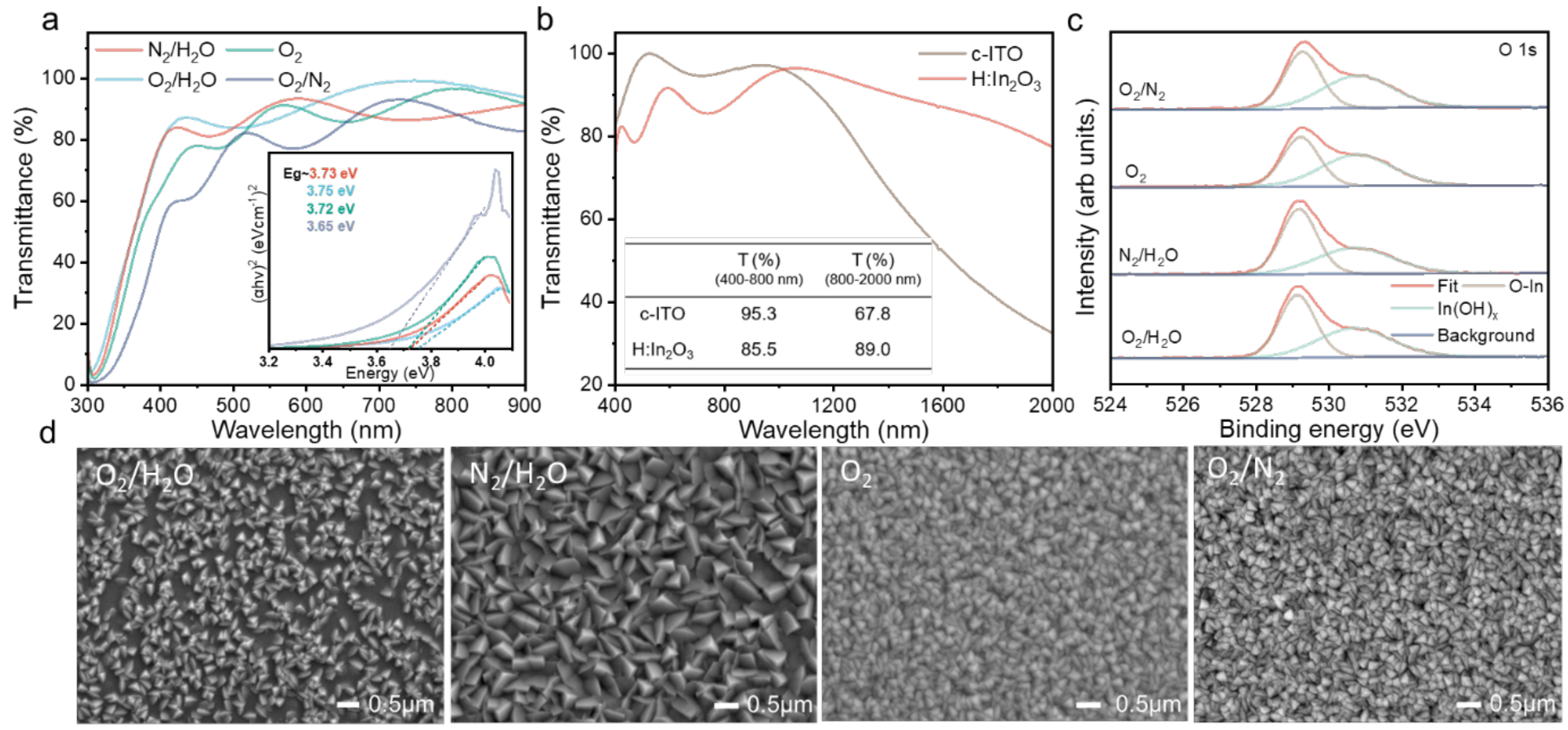

**Figure 2**. Effect of oxidant on optical properties, surface chemistry and morphology of AP-CVD $H:In_2O_3$ films. **a.** Transmittance and bandgap of $H:In_2O_3$ thin films deposited via AP-CVD at 140 °C for 1500 cycles under various oxidizing environments. The thicknesses of film grown with $O_2/H_2O$, $N_2/H_2O$, $O_2$ alone and $O_2/N_2$ in the oxidant channel are 320±30, 470±40, 500±40, 560±50 nm, respectively. **b.** Visible-IR transmittance of commercial ITO (150 ± 3 nm thickness) and optimal $N_2/H_2O$-grown $H:In_2O_3$ thin film (710±50 nm thickness) deposited at 140 °C for 2000 cycles. **c.** XPS spectra, and **d.** high-resolution SEM images of $H:In_2O_3$ thin films deposited under various oxidizing environments for 1500 cycles.

The X-ray photoelectron spectroscopy (XPS) survey spectra of the $In_2O_3$ films (**Figure S5**a, SI) reveal indium and oxygen signals in all samples (details in Supplementary Note 1 and Figure S5a, SI), confirming the formation of $In_2O_3$ [60-62]. The O 1s spectrum reveals two distinct peaks (Figure 2c): $O_I$ at ~529.2 eV, corresponding to In–O–In bonding, and $O_{II}$ at ~530.8 eV, associated with oxygen vacancies [60, 61, 63]. The relative intensity of the $O_{II}$ peak varies across different oxidant environments, indicating different levels of oxygen vacancy concentration [60, 61, 63]. The $N_2/H_2O$-deposited film exhibits the lowest $O_{II}$ peak area, signifying the lowest oxygen vacancy formation, whereas the $O_2$-only film shows the highest $O_{II}$ peak area, indicating a greater density of oxygen vacancy-related defects. The In 3*d* core-level spectra (Figure S5b, SI) further supports this observation, as illustrated in **Table S3**, SI. Although it may seem counterintuitive that films deposited under $O_2$-only conditions exhibited a higher vacancy-related XPS peak, high oxygen partial pressure during AP-CVD can promote rapid precursor combustion and non-equilibrium surface reactions, potentially leading to structural disorder and defect formation. In contrast, using $N_2/H_2O$ in the oxidant channel may be conducive towards more moderate oxidation kinetics, and facilitate controlled hydrogen incorporation, thereby suppressing oxygen vacancy formation and enabling defect passivation. Hydrogen incorporation acts as a shallow donor and can passivate deep oxygen-vacancy–related states, as well as neutralize compensating acceptor-type defects, such as indium vacancies. The relevant defect reactions and energetic considerations are summarized in Supplementary Note 1. These hydrogen-mediated processes could reduce vacancy-related disorder and contribute to improved carrier transport.

Structural characterization further supports this interpretation. XRD patterns (**Figure S4**, SI) confirm the cubic bixbyite phase (space group: $Ia\overline{3}$, No. 206) for all films. The $N_2/H_2O$-grown film exhibits the sharpest and most intense (222) reflection, indicating a higher degree of crystalline ordering. In contrast, $O_2$-only films with similar or even higher thicknesses show broader reflections and mixed orientations, consistent with increased structural disorder. High-resolution SEM images (Figure 2d) reveal corresponding microstructural differences. The $N_2/H_2O$ films exhibit the largest grain size and most compact morphology, whereas $O_2$-containing conditions produce smaller grains and more porous microstructures. Larger grains reduce grain-boundary density and are expected to decrease carrier scattering. These structural trends correlate directly with the electrical data presented in Table 1. The $N_2/H_2O$-grown films, which exhibit the largest grains and highest crystalline ordering, also demonstrate the highest mobility (160±30 $cm^2\ V^{-1}\ s^{-1}$) and lowest resistivity (0.7±0.1mΩ cm). Reduced grain-boundary scattering and lower oxygen-vacancy-related disorder provide a structural and chemical basis for the enhanced transport properties and reduced infrared optical losses using $N_2/H_2O$ [21, 64, 65].

Although figures of merit (FoMs) are often used to benchmark transparent conductors, their values depend strongly on film thickness and the spectral window chosen for averaging transmittance. In this work, FoM calculations (Figure S6-7 & Table S4, SI) indicate that the optimized $H:In_2O_3$ films are comparable to widely-used TCOs when evaluated over the visible wavelength range. However, because conventional FoM formulations typically exclude the NIR region, they do not fully capture the infrared transparency advantage offered by AP-CVD $H:In_2O_3$ (Supplementary Note 2). Accordingly, we emphasize thickness-independent parameters, such as resistivity and mobility when comparing deposition conditions. A broader comparison with literature-reported ITO, $In_2O_3$, and $H:In_2O_3$ films deposited by various techniques is summarized in Table S5, SI. It is important to distinguish between the intrinsic advantages of $H:In_2O_3$ as a material and the advantages of AP-CVD as a deposition method. The high mobility and enhanced near-infrared transmittance originate from the $H:In_2O_3$ system itself, whereas AP-CVD offers a route to realize these properties using low-temperature, atmospheric-pressure processing. In this respect, AP-CVD compares favorably with ALD by delivering similar $H:In_2O_3$ film properties at much higher growth rates. AP-CVD is also favorable compared to sputtering by achieving similarly low resistivities and even higher mobilities without the use of plasma, thereby providing a milder process for sensitive substrates and device stacks.

To elucidate the mechanism of hydrogen incorporation in $H:In_2O_3$, it is essential to distinguish between hydrogen originating from the water oxidant versus atmospheric exposure or residual precursor ligands (*e.g.*, Cp-derived C–H species). To address this, we performed isotope-labelling experiments by replacing $H_2O$ with deuterium oxide ($D_2O$) during deposition, enabling discrimination between hydrogen incorporated from the oxidant source and hydrogen introduced from extrinsic sources such as the atmosphere and precursors. It should be noted that although isotopic substitution from $H_2O$ to $D_2O$ may introduce minor kinetic isotope effects due to differences in O–H and O–D bond energies, the primary objective of these experiments is to identify the origin and depth distribution of hydrogen. The compositional trends observed here therefore support conclusions regarding hydrogen incorporation mechanisms, independent of potential small differences in reaction kinetics.

Nano-secondary ion mass spectrometry (Nano-SIMS) analysis was first conducted to map the depth-dependent deuterium-to-hydrogen (D/H) ratio (**Figure 3**a). As a control experiment, a film deposited under $O_2$-only conditions at 100 °C for 1500 cycles exhibited a D/H ratio of ~0.0005, consistent with the natural isotopic abundance of deuterium in atmospheric hydrogen [66, 67]. Note that we had to reduce the deposition temperature from 140 °C to 100 °C in order to obtain more amorphous films that presented a smoother sample surface for analysis, which is required for SIMS. Our measurements confirm that hydrogen detected using $O_2$-only as the oxidant originates primarily from background contamination or residual Cp-related fragments (Figure S8, SI). In contrast, when the thin film was deposited with $N_2/D_2O$ in the oxidant channel at 100 °C for 1500 cycles, the D/H ratio exhibited a clear depth-dependent trend. Near the film surface, the D/H ratio was approximately 0.2, and increased to approximately 0.25 near the film/substrate interface. This D/H ratio is over 2.5 orders of magnitude greater than the film deposited without $D_2O$, providing strong evidence for the incorporation of hydrogen from the water-based oxidant used.

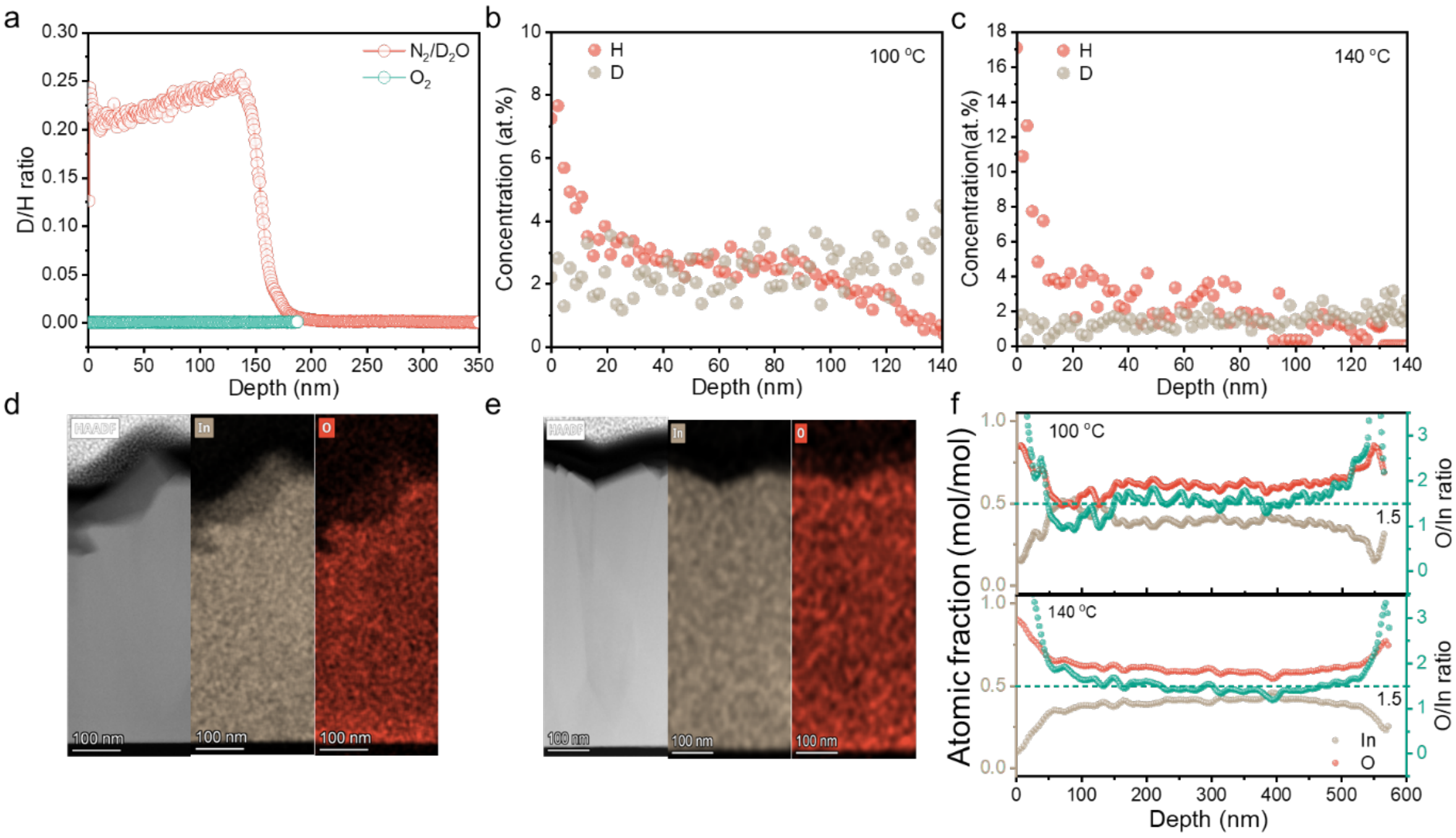

**Figure 3**. Understanding the mechanism of H doping in AP-CVD $H:In_2O_3$. **a.** Depth profiles of the D/H ratio in $H:In_2O_3$ thin films (thickness ~150 nm) deposited onto Si wafer at 100 °C using $N_2/D_2O$ or $O_2$-

only in the oxidant channel, obtained from Nano-SIMS measurements. Depth profiles of D and H atomic concentrations in $H:In_2O_3$ thin films (thickness ~150 nm) deposited onto Si wafer at **b.** 100 °C and **c.** 140 °C using $N_2/D_2O$ in the oxidant channel, obtained from ToF-ERDA measurements. STEM-EDX cross-section images and elemental (In, O) mappings of $H:In_2O_3$ films (thickness ~500 nm) deposited at **d.** 100 °C and **e.** 140 °C. **f.** Depth profiles of the atomic fractions of In and O, along with the O/In ratio, for $H:In_2O_3$ thin films deposited at 100 °C and 140 °C.

To quantify the distribution of hydrogen, we performed time-of-flight elastic recoil detection analysis (ToF-ERDA) measurements on $N_2/D_2O$-grown thin films at both 100 °C and 140 °C, as indicated in Figure 3b-c. While Nano-SIMS provides relative isotopic ratios, ToF-ERDA enables absolute quantification of light elements, allowing direct assessment of hydrogen incorporation and its dependence on deposition temperature [68]. The deuterium concentration remains relatively uniform throughout the film thickness, confirming homogeneous incorporation during growth, supporting a CVD-like hydrogen incorporation mechanism in which water-derived hydrogen is incorporated during each deposition cycle. In contrast, the hydrogen signal decreases near the film surface and exhibits an apparent reduction beyond ~90–100 nm depth. This reduction is attributed to decreased recoil detection efficiency for hydrogen at greater depths, rather than a true absence of hydrogen, as light elements such as hydrogen possess limited recoil energy in ToF-ERDA measurements. After renormalization excluding substrate-derived silicon contributions, the total hydrogen content (H + D) is comparable for both deposition temperatures (~4.8 at.%). This indicates that the overall hydrogen incorporation level is not strongly temperature dependent within the investigated range (**Table S6**, SI). However, the relative contribution of oxidant-derived hydrogen (deuterium) decreases at 140 °C compared to 100 °C. At 100 °C, the D concentration is higher, indicating that hydrogen incorporated from the water oxidant dominates. At 140 °C, although the total hydrogen content in the film remains similar as at 100 °C, the reduced D fraction and increased H fraction suggest a shift in hydrogen incorporation pathway, with a greater relative contribution from metal precursor-derived or background hydrogen sources. This trend is consistent with enhanced precursor decomposition at elevated temperature [69]. Carbon concentration also increases at 140 °C (Figure S9, SI), supporting increased thermal decomposition of the InCp precursor. Carbon-containing fragments may modify the local bonding environment and influence hydrogen chemical state, potentially affecting dopant activation efficiency. Importantly, when considered together with Table 1 and the XPS analysis above, these results suggest that hydrogen originating from the water oxidant is most closely associated with the improved electrical properties. Films grown with water-containing oxidant environments ($O_2/H_2O$ or $N_2/H_2O$) exhibit higher carrier concentrations and lower oxygen-vacancy content, consistent with water-derived hydrogen acting both as a shallow donor and as a passivating species that suppresses oxygen-vacancy-related scattering. This provides a consistent explanation for the simultaneous increase in carrier concentration and conductivity observed for the $H_2O$-grown films.

To further assess the influence of oxidizing environment on overall composition, ToF-ERDA measurements were also performed on films deposited under different oxidants at 140 °C (Table S7, SI). Films grown under $O_2$-only and $O_2/N_2$ conditions exhibit significantly higher carbon contamination (3.3% and 1.9%, respectively) compared to $N_2/H_2O$ and $O_2/H_2O$ films (0.2% and 0.6%). Although elevated temperature promotes precursor fragmentation, the oxidant chemistry determines whether Cp-derived fragments are efficiently removed or retained within the growing film. Under $O_2$-only conditions, rapid surface oxidation may lead to non-equilibrium growth and incomplete ligand removal, resulting in increased carbon incorporation. In contrast, $H_2O$-containing environments appear to facilitate more effective ligand elimination, yielding lower residual carbon content. Carbon impurities are known to introduce mid-gap states that increase sub-bandgap absorption and degrade optical transmittance [69]. The reduced carbon contamination observed for $N_2/H_2O$ and $O_2/H_2O$ films is therefore consistent with their improved transmittance and reduced optical losses (Figure 2a).

The In and O distribution was investigated via Scanning Transmission Electron Microscopy-Energy Dispersive X-ray analysis (STEM-EDX). TEM-EDX elemental mapping of the cross-section of the $H:In_2O_3$ films deposited at 100 and 140 °C shows that both In and O are uniformly distributed across the film thickness, confirming good compositional homogeneity (Figure 3d-e). For the film deposited at 140 °C, the O/In ratio remained close to the stoichiometric value of 1.5 throughout the lamella, with small deviations at the film surface and at the film–substrate interface, which can be attributed to oxygen adsorption from air and the presence of $SiO_x$ at the Si substrate (Figure 3f). Overall, the combined Nano-SIMS, ToF-ERDA, and STEM-EDX analyses demonstrate that hydrogen in AP-CVD-grown $H:In_2O_3$ originates predominantly from the water oxidant, is incorporated uniformly during growth, and is strongly influenced by oxidant chemistry and deposition temperature. These factors collectively govern defect passivation, stoichiometry, and ultimately the electrical and optical properties of the films.

**Conclusions**

In conclusion, we have demonstrated the rapid, atmospheric growth of high figure-of-merit TCOs using AP-CVD. We achieved phase-pure thin films of $H:In_2O_3$ (cubic bixbyite phase, with space group $Ia\overline{3}$) at a deposition temperature of only 140 ºC, without requiring any ozone or plasma. By investigating in detail the effects of the oxidants used, we found that using $H_2O$ vapor with $N_2$ dilution gas resulted in higher-performance $H:In_2O_3$ than using $O_2$ gas as the oxidant. Isotope labelling experiments (using $D_2O$), together with nano-SIMS and ToF-ERDA measurements reveal the benefits of $H_2O$ as the oxidant by introducing H as shallow donor dopants, which increases the carrier concentration, in contrast to the more passivate role of H introduced from the InCp precursor. These H dopants may also passivate oxygen vacancies. This is consistent with XPS measurements that show a lower oxygen vacancy concentration in the films grown with $H_2O$ rather than $O_2$. The mobility of these films improved from 40±10 $cm^2 V^{-1} s^{-1}$ (using $O_2$) to 160±30 $cm^2 V^{-1} s^{-1}$ (using $H_2O$), which is consistent with reduced carrier scattering from oxygen vacancies. Furthermore, by increasing film thicknesses from 470±40 nm to 710±50 nm, we increased mobilities further to 190±30 $cm^2 V^{-1} s^{-1}$. Given the high mobilities, we achieved low sheet resistance down to 7.20±0.01 $\Omega \square^{-1}$ (0.50±0.06 mΩ cm resistivity) without requiring excessively high carrier concentrations, thus avoiding high free carrier absorption. This allowed us to maintain high NIR transmittance (89%) as well as high visible transmittance (85.5%), outperforming commercial ITO (13±1 $\Omega \square^{-1}$ sheet resistance, 67.8% NIR transmittance). Furthermore, despite the mild growth conditions, we achieved 40× higher growth rate than ALD (0.19±0.01 nm $s^{-1}$ vs. ~0.005 nm $s^{-1}$), such that the total processing time required to achieve 0.2–0.8 mΩ cm resistivity $H:In_2O_3$ films was lowered from >160 min (ALD) to only 40±2 min with AP-CVD. This is competitive with sputter deposition of ITO (25±5 min), but with AP-CVD, we have the advantages of fully atmospheric processing (which could enable lower capital intensity of manufacturing) as well as gentler deposition, which is essential for growth onto soft semiconductors. Overall, this work establishes AP-CVD as an appealing method for the scalable manufacture of TCOs that addresses the processability-performance gap of existing deposition methods.

## Acknowledgements

XPS analysis was recorded at the EPSRC National Facility for Photoelectron Spectroscopy (HarwellXPS) ("HarwellXPS", grant no. EP/Y023587/1, EP/Y023609/1, EP/Y023536/1, EP/Y023552/1 and EP/Y023544/1)'. X. G. and R. L. Z. H. thank UK Research and Innovation for funding through a Frontier Grant (no. EP/X029900/1), awarded through the 2021 ERC Starting Grant scheme. A. S. and R. L. Z. H. thank support from St. John's College Oxford for support through the visiting researcher scheme. The authors acknowledge the Agence Nationale de la Recherche (ANR, France) for support via the project ANR-23-715 CE51-0029-01 "INNOVATION". A.S. would like to thank Toulouse INP for the ETI POLYTANCE project and the CNRS for the PROCEDO project. The ToF-ERDA and RBS measurements were supported through the Surrey Ion Beam Centre which is funded by UK EPSRC (EP/X015491/1). The authors thank Vlad Palitsin for help in the ion beam analysis laboratory at the Surrey Ion Beam Centre. E. L. Q. acknowledges funding from the EPSRC Centre for Doctoral Training in Inorganic Chemistry for Future Manufacturing (OxICFM; no. EP/S023828/1). C.-Y. C. thanks the Oxford-Taiwan Graduate Scholarship and Clarendon Fund Scholarships. J. O. acknowledges funding via the EPSRC Doctoral Prize [Grant No. EP/W524311/1] and the EPSRC Impact Acceleration Account [Grant No. EP/X525777/1]. R. S. B was supported by the Leverhulme Trust Prize (grant RPG-2020-377) and the UK Engineering and Physical Sciences Research Council (grant number EP/X037169/1). H-J. S thanks the National Research Foundation of Korea (NRF) grant funded by the Korea government (MSIT) (RS-2024-00359374). R. L. Z. H. thanks the Royal Academy of Engineering and Science & Technology Facilities Council for support through a Senior Research Fellowship (no. RCSRF/2324-18-68). C. D. acknowledges the support of the Wolfson Electron Microscopy Suite and the use of the Thermo Fisher Spectra 300 TEM, funded by EPSRC grant EP/R008779/1. The authors thank *NanoPrint Innovations Ltd.* and Zhuotong Sun for providing the InCp precursor.

**Author contributions**

X. G., and R. L. Z. H. conceived of the project. X. G. deposited all $H:In_2O_3$ thin films using atmospheric pressure chemical vapor deposition and analyzed all data collected. X. G. measured optical and electrical properties with the assistance of J. O., who is under the supervision of R. S. B. H-J. S. and H-K. K. helped with ellipsometer measurement. E. L. Q., M. K. S. and C. D. M. conducted Time-of-flight elastic recoil detection analysis. Y-T. H. helped with the scanning electron microscope measurement. X. L. conducted scanning transmission electron microscopy- energy dispersive X-ray analysis under the supervision of C. D. K. L. conducted Nano-SIMS analysis under the supervision of K. L. M. C.-Y. C. and Y. W. contributed to discussions of potential applications of the thin films. A. S. contributed to conceive the project. R. L. Z. H. supervised the overall work. All authors contributed to the writing and editing of the manuscript.

**Conflicts of interest**

R. L. Z. H. declares a competing financial interest, in that he is Chief Technology Officer of *NanoPrint Innovations Ltd.*, a company manufacturing spatial atomic layer deposition equipment.

**Reference**

1. Althumayri, M.; Das, R.; Banavath, R.; Beker, L.; Achim, A. M.; Ceylan Koydemir, H., "Recent advances in transparent electrodes and their multimodal sensing applications," *Advanced Science* 11, no. 38 (2024): 2405099.
2. Lee, H. B.; Jin, W.-Y.; Ovhal, M. M.; Kumar, N.; Kang, J.-W., "Flexible transparent conducting electrodes based on metal meshes for organic optoelectronic device applications: a review," *Journal of Materials Chemistry C* 7, no. 5 (2019): 1087-1110.
3. Li, J.; Liang, J.; Li, L.; Ren, F.; Hu, W.; Li, J.; Qi, S.; Pei, Q., "Healable capacitive touch screen sensors based on transparent composite electrodes comprising silver nanowires and a furan/maleimide Diels–Alder cycloaddition polymer," *ACS nano* 8, no. 12 (2014): 12874-12882.
4. Guo, X.; Chen, J.; Eh, A. L.-S.; Poh, W. C.; Jiang, F.; Jiang, F.; Chen, J.; Lee, P. S., "Heat-insulating black electrochromic device enabled by reversible nickel–copper electrodeposition," *ACS Applied Materials & Interfaces* 14, no. 17 (2022): 20237-20246.
5. Sekkat, A.; Sanchez-Velasquez, C.; Bardet, L.; Weber, M.; Jiménez, C.; Bellet, D.; Muñoz-Rojas, D.; Nguyen, V. H., "Towards enhanced transparent conductive nanocomposites based on metallic nanowire networks coated with metal oxides: a brief review," *Journal of Materials Chemistry A* 12, no. 38 (2024): 25600-25621.
6. Espid, E.; Taghipour, F., "Development of highly sensitive ZnO/In2O3 composite gas sensor activated by UV-LED," *Sensors and Actuators B: Chemical* 241, no. (2017): 828-839.
7. Guo, X.; Chen, S.; Poh, W. C.; Chen, J.; Jiang, F.; Tan, M. W. M.; Lee, P. S., "Solid-State and Flexible Black Electrochromic Devices Enabled by Ni-Cu Salts Based Organohydrogel Electrolytes," *Advanced Materials Interfaces* 10, no. 15 (2023): 2300061.
8. Koida, T., "Amorphous and crystalline In2O3-based transparent conducting films for photovoltaics," *physica status solidi (a)* 214, no. 2 (2017): 1600464.
9. Wang, C. Y.; Cimalla, V.; Kups, T.; Röhlig, C.-C.; Stauden, T.; Ambacher, O.; Kunzer, M.; Passow, T.; Schirmacher, W.; Pletschen, W., "Integration of In2O3 nanoparticle based ozone sensors with GaInN∕GaN light emitting diodes," *Applied Physics Letters* 91, no. 10 (2007):
10. Johnson, K.; Guruvenket, S.; Jha, S.; Halverson, B.; Olson, C.; Sailer, R.; Pokhodnya, K.; Schulz, D. In *Atmospheric-pressure plasma deposition of indium tin oxide*, 2009 34th IEEE Photovoltaic Specialists Conference (PVSC), IEEE: **2009**; pp 001806-001810.
11. Gaskell, J. M.; Sheel, D. W., "Deposition of indium tin oxide by atmospheric pressure chemical vapour deposition," *Thin Solid Films* 520, no. 12 (2012): 4110-4113.
12. Suzuki, R.; Nishi, Y.; Matsubara, M.; Muramatsu, A.; Kanie, K., "A nanoparticle-mist deposition method: Fabrication of high-performance ITO flexible thin films under atmospheric conditions," *Scientific Reports* 11, no. 1 (2021): 10584.
13. Cho, H.; Yun, Y.-H., "Characterization of indium tin oxide (ITO) thin films prepared by a sol–gel spin coating process," *Ceramics International* 37, no. 2 (2011): 615-619.
14. Peelaers, H.; Kioupakis, E.; Van de Walle, C. G., "Fundamental limits on optical transparency of transparent conducting oxides: Free-carrier absorption in SnO2," *Applied Physics Letters* 100, no. 1 (2012):
15. Egbo, K. O.; Adesina, A. E.; Ezeh, C. V.; Liu, C. P.; Yu, K. M., "Effects of free carriers on the optical properties of high mobility transition metal doped In 2 O 3 transparent conductors," *Physical Review Materials* 5, no. 9 (2021): 094603.
16. Tan, L.; Fu, Y.; Kang, S.; Wondraczek, L.; Lin, C.; Yue, Y., "Broadband NIR-emitting Te cluster-doped glass for smart light source towards night-vision and NIR spectroscopy applications," *Photonics Research* 10, no. 5 (2022): 1187-1193.
17. Wang, Z.; Chen, C.; Wu, K.; Chong, H.; Ye, H., "Transparent conductive oxides and their applications in near infrared plasmonics," *physica status solidi (a)* 216, no. 5 (2019): 1700794.
18. Koida, T.; Fujiwara, H.; Kondo, M., "Hydrogen-doped In2O3 as high-mobility transparent conductive oxide," *Japanese Journal of Applied Physics* 46, no. 7L (2007): L685.
19. Macco, B.; Wu, Y.; Vanhemel, D.; Kessels, W., "High mobility In2O3: H transparent conductive oxides prepared by atomic layer deposition and solid phase crystallization," *physica status solidi (RRL)–Rapid Research Letters* 8, no. 12 (2014): 987-990.
20. Macco, B.; Knoops, H. C.; Kessels, W. M., "Electron scattering and doping mechanisms in solid-phase-crystallized In2O3: H prepared by atomic layer deposition," *ACS applied materials & interfaces* 7, no. 30 (2015): 16723-16729.
21. Magari, Y.; Kataoka, T.; Yeh, W.; Furuta, M., "High-mobility hydrogenated polycrystalline In2O3 (In2O3: H) thin-film transistors," *Nature communications* 13, no. 1 (2022): 1078.
22. Ge, C.; Liu, Z.; Zhu, Y.; Zhou, Y.; Jiang, B.; Zhu, J.; Yang, X.; Zhu, Y.; Yan, S.; Hu, H., "Insight into the high mobility and stability of In2O3: H film," *Small* 20, no. 2 (2024): 2304721.
23. Barraud, L.; Holman, Z.; Badel, N.; Reiss, P.; Descoeudres, A.; Battaglia, C.; De Wolf, S.; Ballif, C., "Hydrogen-doped indium oxide/indium tin oxide bilayers for high-efficiency silicon heterojunction solar cells," *Solar Energy Materials and Solar Cells* 115, no. (2013): 151-156.
24. Park, H. G.; Hussain, S. Q.; Park, J.; Yi, J., "Influence of hydrogen doping of In2O3-based transparent conducting oxide films on silicon heterojunction solar cells," *Journal of Materials Science* 59, no. 30 (2024): 13873-13882.
25. Koida, T.; Ueno, Y.; Shibata, H., "In2O3-based transparent conducting oxide films with high electron mobility fabricated at low process temperatures," *physica status solidi (a)* 215, no. 7 (2018): 1700506.

26. Mameli, A.; Kuang, Y.; Aghaee, M.; Ande, C. K.; Karasulu, B.; Creatore, M.; Mackus, A. J.; Kessels, W. M.; Roozeboom, F., "Area-selective atomic layer deposition of In2O3: H using a μ-plasma printer for local area activation," *Chemistry of Materials* 29, no. 3 (2017): 921-925.
27. Wu, Y.; Macco, B.; Vanhemel, D.; Kolling, S.; Verheijen, M. A.; Koenraad, P. M.; Kessels, W. M.; Roozeboom, F., "Atomic layer deposition of In2O3: H from InCp and H2O/O2: microstructure and isotope labeling studies," *ACS applied materials & interfaces* 9, no. 1 (2017): 592-601.
28. Kim, H. Y.; Jung, E. A.; Mun, G.; Agbenyeke, R. E.; Park, B. K.; Park, J.-S.; Son, S. U.; Jeon, D. J.; Park, S.-H. K.; Chung, T.-M., "Low-temperature growth of indium oxide thin film by plasma-enhanced atomic layer deposition using liquid dimethyl (N-ethoxy-2, 2-dimethylpropanamido) indium for high-mobility thin film transistor application," *ACS Applied Materials & Interfaces* 8, no. 40 (2016): 26924-26931.
29. Asikainen, T.; Ritala, M.; Leskelä, M., "Growth of In2S3 thin films by atomic layer epitaxy," *Applied Surface Science* 82, no. (1994): 122-125.
30. Putkonen, M.; Sajavaara, T.; Niinistö, L., "Enhanced growth rate in atomic layer epitaxy deposition of magnesium oxide thin films," *Journal of Materials Chemistry* 10, no. 8 (2000): 1857-1861.
31. Ritala, M.; Asikainen, T.; Leskelä, M.; Skarp, J., "Ale growth of transparent conductors," *MRS Online Proceedings Library* 426, no. (1996): 513-518.
32. Mizutani, F.; Higashi, S.; Inoue, M.; Nabatame, T., "Atomic layer deposition of stoichiometric In2O3 films using liquid ethylcyclopentadienyl indium and combinations of H2O and O2 plasma," *AIP Advances* 9, no. 4 (2019):
33. Ma, Q.; Zheng, H.-M.; Shao, Y.; Zhu, B.; Liu, W.-J.; Ding, S.-J.; Zhang, D. W., "Atomic-layer-deposition of indium oxide nano-films for thin-film transistors," *Nanoscale research letters* 13, no. (2018): 1-8.
34. Agbenyeke, R. E.; Jung, E. A.; Park, B. K.; Chung, T.-M.; Kim, C. G.; Han, J. H., "Thermal atomic layer deposition of In2O3 thin films using dimethyl (N-ethoxy-2, 2-dimethylcarboxylicpropanamide) indium and H2O," *Applied Surface Science* 419, no. (2017): 758-763.
35. Ma, Q.; Shao, Y.; Wang, Y.-P.; Zheng, H.-M.; Zhu, B.; Liu, W.-J.; Ding, S.-J.; Zhang, D., "Rapid improvement in thin film transistors with atomic-layer-deposited InO x channels via O 2 plasma treatment," *IEEE Electron Device Letters* 39, no. 11 (2018): 1672-1675.
36. Libera, J. A.; Hryn, J. N.; Elam, J. W., "Indium oxide atomic layer deposition facilitated by the synergy between oxygen and water," *Chemistry of Materials* 23, no. 8 (2011): 2150-2158.
37. Zhao, M.-J.; Zhang, Z.-X.; Hsu, C.-H.; Zhang, X.-Y.; Wu, W.-Y.; Lien, S.-Y.; Zhu, W.-Z., "Properties and mechanism of PEALD-In2O3 thin films prepared by different precursor reaction energy," *Nanomaterials* 11, no. 4 (2021): 978.
38. Elam, J. W.; Martinson, A. B.; Pellin, M. J.; Hupp, J. T., "Atomic layer deposition of In2O3 using cyclopentadienyl indium: a new synthetic route to transparent conducting oxide films," *Chemistry of Materials* 18, no. 15 (2006): 3571-3578.
39. Elam, J. W.; Libera, J. A.; Hryn, J. N., "Indium oxide ALD using Cyclopentadienyl indium and mixtures of H2O and O2," *Ecs Transactions* 41, no. 2 (2011): 147.
40. Hoye, R. L.; Muñoz-Rojas, D.; Sun, Z.; Okcu, H.; Asgarimoghaddam, H.; MacManus-Driscoll, J. L.; Musselman, K. P., "Spatial atomic layer deposition for energy and electronic devices," *PRX Energy* 4, no. 1 (2025): 017002.
41. Levy, D. H.; Freeman, D.; Nelson, S. F.; Cowdery-Corvan, P. J.; Irving, L. M., "Stable ZnO thin film transistors by fast open air atomic layer deposition," *Applied Physics Letters* 92, no. 19 (2008):
42. Poodt, P.; Lankhorst, A.; Roozeboom, F.; Spee, K.; Maas, D.; Vermeer, A., "High-speed spatial atomic-layer deposition of aluminum oxide layers for solar cell passivation," *Adv. Mater* 22, no. 32 (2010): 3564-3567.
43. Levy, D. H.; Nelson, S. F., "Thin-film electronics by atomic layer deposition," *Journal of Vacuum Science & Technology A* 30, no. 1 (2012):
44. Muñoz-Rojas, D.; Jordan, M.; Yeoh, C.; Marin, A.; Kursumovic, A.; Dunlop, L.; Iza, D.; Chen, A.; Wang, H.; MacManus Driscoll, J., "Growth of∼ 5 cm2V− 1s− 1 mobility, p-type Copper (I) oxide (Cu2O) films by fast atmospheric atomic layer deposition (AALD) at 225 C and below," *AIP Advances* 2, no. 4 (2012):

45. Muñoz-Rojas, D.; Sun, H.; Iza, D. C.; Weickert, J.; Chen, L.; Wang, H.; Schmidt-Mende, L.; MacManus-Driscoll, J. L., "High-speed atmospheric atomic layer deposition of ultra thin amorphous TiO2 blocking layers at 100 C for inverted bulk heterojunction solar cells," *Progress in Photovoltaics: Research and Applications* 21, no. 4 (2013): 393-400.
46. Hoye, R. L.; Musselman, K. P.; MacManus-Driscoll, J. L., "Research update: Doping ZnO and TiO2 for solar cells," *APL materials* 1, no. 6 (2013):
47. Marin, A. T.; Muñoz-Rojas, D.; Iza, D. C.; Gershon, T.; Musselman, K. P.; MacManus-Driscoll, J. L., "Novel atmospheric growth technique to improve both light absorption and charge collection in ZnO/Cu2O thin film solar cells," *Advanced Functional Materials* 23, no. 27 (2013): 3413-3419.
48. Munoz-Rojas, D.; MacManus-Driscoll, J., "Spatial atmospheric atomic layer deposition: a new laboratory and industrial tool for low-cost photovoltaics," *Materials Horizons* 1, no. 3 (2014): 314-320.
49. Yoo, K. S.; Lee, C.-H.; Kim, D.-G.; Choi, S.-H.; Lee, W.-B.; Park, C.-K.; Park, J.-S., "High mobility and productivity of flexible In2O3 thin-film transistors on polyimide substrates via atmospheric pressure spatial atomic layer deposition," *Applied Surface Science* 646, no. (2024): 158950.
50. Hoye, R. L.; Muñoz-Rojas, D.; Musselman, K. P.; Vaynzof, Y.; MacManus-Driscoll, J. L., "Synthesis and modeling of uniform complex metal oxides by close-proximity atmospheric pressure chemical vapor deposition," *ACS Applied Materials & Interfaces* 7, no. 20 (2015): 10684-10694.
51. Jagt, R. A.; Huq, T. N.; Hill, S. A.; Thway, M.; Liu, T.; Napari, M.; Roose, B.; Gałkowski, K.; Li, W.; Lin, S. F., "Rapid vapor-phase deposition of high-mobility p-type buffer layers on perovskite photovoltaics for efficient semitransparent devices," *ACS Energy Letters* 5, no. 8 (2020): 2456-2465.
52. O'Sullivan, J.; Wright, M.; Stefani, B. V.; Llontop, P.; Sekkat, A.; Hoye, R. L. Z.; Morales-Masis, M.; Bonilla, R. S., "Transparent conducting electrodes for perovskite-silicon tandem solar cells," *nature reviews physics* no. (2026): https://doi.org/https://doi.org/10.1038/s42254-026-00938-5
53. Hoye, R. L.; Muñoz-Rojas, D.; Nelson, S. F.; Illiberi, A.; Poodt, P.; Roozeboom, F.; MacManus-Driscoll, J. L., "Research Update: Atmospheric pressure spatial atomic layer deposition of ZnO thin films: Reactors, doping, and devices," *APL materials* 3, no. 4 (2015):
54. Tuna, O.; Selamet, Y.; Aygun, G.; Ozyuzer, L., "High quality ITO thin films grown by dc and RF sputtering without oxygen," *Journal of Physics D: Applied Physics* 43, no. 5 (2010): 055402.
55. Wang, K.; Jiao, P.; Cheng, Y.; Xu, H.; Zhu, G.; Zhao, Y.; Jiang, K.; Zhang, X.; Su, Y., "ITO films with different preferred orientations prepared by DC magnetron sputtering," *Optical Materials* 134, no. (2022): 113040.
56. Park, J.; Buurma, C.; Sivananthan, S.; Kodama, R.; Gao, W.; Gessert, T., "The effect of post-annealing on Indium Tin Oxide thin films by magnetron sputtering method," *Applied Surface Science* 307, no. (2014): 388-392.
57. Senol, S.; Senol, A.; Ozturk, O.; Erdem, M., "Effect of annealing time on the structural, optical and electrical characteristics of DC sputtered ITO thin films," *Journal of Materials Science: Materials in Electronics* 25, no. 11 (2014): 4992-4999.
58. Raninga, R. D.; Jagt, R. A.; Béchu, S.; Huq, T. N.; Li, W.; Nikolka, M.; Lin, Y.-H.; Sun, M.; Li, Z.; Li, W., "Strong performance enhancement in lead-halide perovskite solar cells through rapid, atmospheric deposition of n-type buffer layer oxides," *Nano Energy* 75, no. (2020): 104946.
59. He, H.; Shetty, N.; Bauch, T.; Kubatkin, S.; Kaufmann, T.; Cornils, M.; Yakimova, R.; Lara-Avila, S., "The performance limits of epigraphene Hall sensors doped across the Dirac point," *Applied Physics Letters* 116, no. 22 (2020):
60. Gan, J.; Lu, X.; Wu, J.; Xie, S.; Zhai, T.; Yu, M.; Zhang, Z.; Mao, Y.; Wang, S. C. I.; Shen, Y., "Oxygen vacancies promoting photoelectrochemical performance of In2O3 nanocubes," *Scientific reports* 3, no. 1 (2013): 1021.
61. Sun, L.; Li, R.; Zhan, W.; Yuan, Y.; Wang, X.; Han, X.; Zhao, Y., "Double-shelled hollow rods assembled from nitrogen/sulfur-codoped carbon coated indium oxide nanoparticles as excellent photocatalysts," *Nature Communications* 10, no. 1 (2019): 2270.
62. Pan, B.; Yuan, G.; Zhao, X.; Han, N.; Huang, Y.; Feng, K.; Cheng, C.; Zhong, J.; Zhang, L.; Wang, Y., "Highly dispersed indium oxide nanoparticles supported on carbon nanorods enabling efficient electrochemical CO2 reduction," *Small Science* 1, no. 10 (2021): 2100029.

63. Kyndiah, A.; Ablat, A.; Guyot-Reeb, S.; Schultz, T.; Zu, F.; Koch, N.; Amsalem, P.; Chiodini, S.; Yilmaz Alic, T.; Topal, Y., "A multifunctional interlayer for solution processed high performance indium oxide transistors," *Scientific Reports* 8, no. 1 (2018): 10946.
64. Wardenga, H. F.; Frischbier, M. V.; Morales-Masis, M.; Klein, A., "In situ hall effect monitoring of vacuum annealing of In2O3: H thin films," *Materials* 8, no. 2 (2015): 561-574.
65. Muydinov, R.; Steigert, A.; Wollgarten, M.; Michałowski, P. P.; Bloeck, U.; Pflug, A.; Erfurt, D.; Klenk, R.; Körner, S.; Lauermann, I., "Crystallisation phenomena of In2O3: H films," *Materials* 12, no. 2 (2019): 266.
66. O'meara, J. M.; Tytler, D.; Kirkman, D.; Suzuki, N.; Prochaska, J. X.; Lubin, D.; Wolfe, A. M., "The Deuterium to Hydrogen Abundance Ratio toward a Fourth QSO: HS0105+ 1619," *The Astrophysical Journal* 552, no. 2 (2001): 718.
67. Kirkman, D.; Tytler, D.; Suzuki, N.; O'Meara, J. M.; Lubin, D., "The cosmological baryon density from the deuterium-to-hydrogen ratio in QSO absorption systems: D/H toward Q1243+ 3047," *The Astrophysical Journal Supplement Series* 149, no. 1 (2003): 1.
68. Sharpe, M. K.; Wright, M.; McAleese, C. D.; Wang, Y.; Hobson, T. D.; Niu, X.; Morisset, A.; Bonilla, R. S., "Characterisation of solar cell nanolayer interfaces using time-of-flight elastic recoil detection analysis," *Solar energy materials and solar cells* 296, no. (2026): 114074.
69. Wu, Y.; Macco, B.; Vanhemel, D.; Kölling, S.; Verheijen, M. A.; Koenraad, P. M.; Kessels, W. M. M.; Roozeboom, F., "Atomic Layer Deposition of In2O3:H from InCp and H2O/O2: Microstructure and Isotope Labeling Studies," *ACS Applied Materials & Interfaces* 9, no. 1 (2017): 592-601. https://doi.org/10.1021/acsami.6b13560